# i-*R*-Cd (*R* = Gd – Tm, Y): A new family of binary magnetic icosahedral quasicrystals

Alan I. Goldman[1,2,*], Tai Kong[1,2], Andreas Kreyssig[1,2], Anton Jesche[1,2], Mehmet Ramazanoglu[1,2], Kevin W. Dennis[1], Sergey L. Bud'ko[1,2] and Paul C. Canfield[1,2,**]

[1]Ames Laboratory, US DOE, Iowa State University, Ames, Iowa 50011, USA
[2]Department of Physics and Astronomy, Iowa State University, Ames, Iowa 50011, USA

**"Seek and ye shall find, the unsought shall go undetected." This adage, attributed to Aristophanes, can be considered one of the defining mantras of new materials research; if you don't look, you certainly will not discover [1]. As a result of our recent discovery of the binary quasicrystalline phase i-$Sc_{12}Zn_{88}$ [2] we proposed that there may well be other binary quasicrystalline phases lurking nearby known crystalline approximants, perhaps as peritectally forming compounds with very limited liquidus surfaces, offering greatly reduced ranges of composition/temperature for primary solidification. Here we report that, as the adage goes, *we have found*. Indeed, adjacent to the *R*$Cd_6$, cubic approximate structure, we have discovered the long sought after model system for magnetic quasicrystals: a new family of at least seven rare earth icosahedral quasicrystals: i-*R*-Cd for *R* = Gd – Tm,Y, six of which now form the known set of moment bearing, binary quasicrystals.**

In support of our discovery of i-*R*-Cd we have experimentally re-evaluated the Cd-rich regions of the binary *R*-Cd phase diagrams and determined the compositions of the deeply peritectic i-*R*-Cd phases. We further show that these magnetic quasicrystals contain the same basic structural elements as found in the non-moment-bearing i-$YbCd_{5.7}$ icosahedral phase [3, 4] and $R$Cd$_6$ periodic approximants despite differences in composition. Finally, we present data that strongly suggests that the new quasicrystal phases enter into a low-temperature spin-glass (SG) state rather than manifest antiferromagnetic order as found for the $R$Cd$_6$ approximants [5-7]. The new i-$R$-Cd system will play a key role as the simplest magnetic quasicrystal system, offering non-magnetic, Y, Heisenberg-like, Gd, and non-Heisenberg (i.e. CEF split) Tb through Tm members, as well as the structural and compositional simplicity of a binary phase.

To date, all of the known quasicrystals with moment bearing elements exhibit frustration and spin-glass-like behavior at low temperatures [8, 9]. In the ternary $R$-Mg-Zn icosahedral quasicrystal, for example, dc- and ac-susceptibility measurements demonstrate spin-glass behavior [10], and neutron diffraction measurements on both $R$-Mg-Zn and $R$-Mg-Cd clearly show the presence of only short-range magnetic correlations at low temperature [11,12]. Interestingly, the absence of long-range magnetic order also extended to the known crystalline approximant phases as well [13-15]. Crystalline approximants are periodic crystals with compositions and unit cell atomic decorations (e.g. atomic clusters) that are closely related to their respective quasicrystalline phases [16]. The $R$Cd$_6$ crystalline phases, which are isostructural to the approximant (YbCd$_6$) of the non-moment-bearing binary i-YbCd$_{5.7}$ icosahedral phase, provide an exception to this trend since both thermodynamic and scattering measurements have demonstrated the onset of long-range antiferromagnetic (AFM) order at low temperatures [5-7]. The observation of AFM order in the $R$Cd$_6$ approximants provided a strong impetus to search for and study a related moment-bearing binary icosahedral phase to better understand, or place constraints on, possibilities for the existence of AFM order on an aperiodic lattice [17].

The new i-R-Cd icosahedral family ($R$ = Gd – Tm, Y) was discovered by using solution growth of single crystals [18,19] as an exploratory synthetic tool. In this case, based on our suspicion that there may be a Cd-rich, quasicrystal phase related to the $R$Cd$_6$ approximants, we slowly cooled melts of $R_{0.07}$Cd$_{0.93}$ from 700 ºC down to 340 ºC and decanted off the remaining (over 99% Cd rich) solution. These first growths revealed large, well developed crystals of $R$Cd$_6$ with clear, second phase, pentagonal dodecahedra, characteristic of solution grown quasicrystals [19], appearing both on the $R$Cd$_6$ phase as well as on the crucible walls (Figure 1 inset). Given that the existing $R$-Cd binary phase diagrams were clearly incomplete on the Cd-rich side of $R$Cd$_6$ we determined the liquidus line for $R$Cd$_6$, the peritectic temperature for the quasicrystalline phase, and set limits on the eutectic composition (Figure 1). With these data we have been able to grow single phase i-$R$-Cd by, for example, slowly cooling melts of Gd$_{0.008}$Cd$_{0.992}$ from 455 ºC to 335 ºC over 50 hours and then decanting off the excess Cd. Even though the exposed liquidus line for i-Gd-Cd formation is small, it shrinks further as we progress from $R$ = Gd to $R$ = Tm; for i-Tm-Cd we could only grow single phase samples from an initial melt of Tm$_{0.006}$Cd$_{0.994}$.

The composition of the icosahedral phase was inferred from two independent methods: wavelength dispersive spectroscopy (WDS) and temperature dependent magnetization (shown in the inset to Figure 2). Values from each measurement are presented with other magnetic and structural data in Table 1. From the WDS measurements we find an average composition of i-$R$Cd$_{7.55\pm0.3}$ across the series whereas, from our fits to the high-temperature $H/M(T)$ data for the moment-bearing samples, we find that the composition is slightly more Cd rich: i-$R$Cd$_{7.75 \pm 0.25}$. Both measurements, however, set compositions for i-$R$-Cd that (i) differ significantly from the prototypical YbCd$_{5.7}$ icosahedral quasicrystal and $R$Cd$_6$ cubic approximants and (ii) are close to the stoichiometry of the recently discovered Sc$_{12}$Zn$_{88}$ (ScZn$_{7.33}$ in the present notation) icosahedral phase [2] as discussed in detail below. We further note that both measurements suggest the possibility of a slight increase in $R$ content for the heavier rare earths relative to $R$ = Gd, which could be driven by steric (ionic size) constraints.

X-ray powder diffraction, using a conventional laboratory source, and high-energy x-ray single-grain diffraction [21], using station 6-ID-D at the Advanced Photon Source, were employed to characterize and index the diffraction patterns from several i-$R$-Cd samples. The single-grain precession images of i-Gd-Cd, shown in Figures 3(a) and (b), were taken with the beam along a fivefold and twofold axis of the pentagonal facetted grain and display the pattern of diffraction spots characteristic of a primitive (P-type) quasilattice. This was also confirmed for $R$ = Tb and Tm. Likewise, the diffraction pattern from powdered single grains of i-Gd-Cd, shown in Figure 3(c), is well-indexed by a primitive icosahedral quasilattice with a 6D lattice constant, $a_{6D}$ = 7.972(4) Å. No impurity phases, beyond some residual Cd flux at the <5% level, were found. As the inset to Figure 3(c) shows, our powder diffraction measurements for $R$ = Gd, Tb, Dy, Ho, Er and Tm show that $a_{6D}$ decreases smoothly, consistent with the well-known lanthanide contraction.

The difference in composition between the new i-$R$-Cd phase and the i-YbCd$_{5.7}$ quasicrystal, as well as the $R$Cd$_6$ quasicrystal approximants, raises the question of whether there are fundamental differences between the structures of these systems. Our analysis of the structure of i-$R$-Cd relative to the i-YbCd$_{5.7}$ quasicrystal and $R$Cd$_6$ approximants is based on the close association between the atomic motifs in quasicrystals and their associated approximants [16]. The $R$Cd$_6$ cubic approximants may be described as a body-centered cubic (bcc) packing of interpenetrating rhombic triacontahedral (RT), or "Tsai-type" clusters [3], which features an icosahedron of 12 $R$ atoms comprising the third shell of the cluster. These clusters, situated at the bcc lattice points are linked along the cubic axes by sharing a face, and interpenetrate neighboring clusters along the body diagonal [24].

These same clusters have been shown to comprise the backbone of the structure of the icosahedral phase of i-YbCd$_{5.7}$, with the same type of linkages [4]. Indeed, in terms of the higher dimensional description of aperiodic crystals, the atomic structure of the $R$Cd$_6$ approximant can be generated by a rational projection of the 6D representation of the icosahedral phase [4]. However, there is only one crystallographic site for the $R$ ion in the approximant structure corresponding to their placement at the vertices of an icosahedron embedded in the RT cluster. For i-YbCd$_{5.7}$, on the other hand, approximately 70% of the $R$ ions are associated with the embedded icosahedral cluster, whereas 30% are found in the "glue" that fills the gaps between the RT clusters [4, 25]. We propose, as previously suggested [26] to explain the low concentration of Sc in i-Sc$_{12}$Zn$_{88}$ [2], that the RT clusters in i-$R$-Cd remain intact but the "glue" filling the gaps between the RT clusters is deficient in $R$ ions (e.g. the complete absence of $R$ ions in the glue would lead to a composition of $R$Cd$_{8.6}$).

In support of this proposal, we compare our measured valued for $a_{6D}$ to the lattice constant, $a$, expected for the $R\text{Cd}_6$ approximant phase through the well-established relation [15]: $a = \dfrac{\sqrt{2}a_{6D}(p+q\tau)}{\sqrt{(2+\tau)}}$, where $\tau$ is the golden mean, ($\dfrac{\sqrt{5}+1}{2}$), and $p$ and $q$ are indices that label the approximant. For the case at hand, $p = q = 1$. For i-Gd-Cd, we find a calculated value of $a = 15.519(6)$ Å, in excellent agreement with our measured lattice parameter for $\text{GdCd}_6$ of $15.523(5)$ Å. As the inset to Figure 3(c) shows, excellent agreement between the derived and measured cubic approximant lattice constants is found for all $R$ ions in the i-$R$-Cd family. This provides compelling evidence that, despite rather significant differences in composition, the cluster-based backbone of the icosahedral structure of the i-$R$-Cd quasicrystal is quite similar, if not identical, to that found for i-$\text{YbCd}_{5.7}$ and the $R\text{Cd}_6$ cubic approximants. Full confirmation of this point will require a full 6D structural refinement of the i-$R$-Cd quasicrystal as was done for i-$\text{YbCd}_{5.7}$ [4].

With the basic structural framework of clusters and linkages in *i-R*-Cd intact we now turn to the detailed magnetic properties of the i-$R$-Cd quasicrystals, particularly in comparison with their AFM ordered $R\text{Cd}_6$ cubic approximants. We begin by returning to the high-temperature $H/M(T)$ measurements shown in Fig. 2. From these data, the Weiss temperature, $\Theta$, is shown together with values previously measured for other ternary rare-earth containing quasicrystalline compounds including i-$R_9\text{Mg}_{34}\text{Zn}_{57}$ [10], i-$R_{10}\text{Mg}_{40}\text{Cd}_{50}$ [11] and i-$\text{Gd}_{14}\text{Ag}_{50}\text{In}_{36}$ [14]. For all of these compounds, $\Theta$ is negative, denoting predominantly AFM exchange interactions, and roughly scales with the de Gennes factor, as expected. What is most striking, however, is that the values for $\Theta$ are very consistent from system-to-system even though these compounds manifest different compositions and structural classes. i-$R_9$Mg$_{34}$Zn$_{57}$, for example, is a ternary face-centered (F-type) icosahedral quasicrystal characterized by Bergman-type clusters, whereas we have now shown that i-$R$-Cd is a binary primitive (P-type) icosahedral quasicrystal characterized by RT clusters. Nevertheless, the strength of the AFM exchange, characterized by $\Theta$, and its monotonic dependence on the de Gennes factor, remains remarkably similar across the spectrum of structures. Although this does not necessarily follow from the aperiodicity of these structures, and may simply be a consequence of the presence in all cases of a pseudo-gap in the density-of-states at the Fermi energy associated with the Hume-Rothery mechanism for the stabilization of quasicrystals [27], this point deserves further study.

To further explore the low-temperature magnetic state of i-$R$-Cd we show the field-cooled (FC) and zero-field-cooled (ZFC) $M(T)/H$ data for i-Gd-Cd, i-Tb-Cd and i-Dy-Cd in Figure 4. These data suggest that the quasicrystalline phase, for $R$ = Gd, Tb and Dy, enters into a low-temperature spin-glass state. i-Gd-Cd exhibits canonical spin-glass behavior where there is a pronounced difference between the FC and ZFC data below a peak which can be cleanly identified with a spin-freezing temperature [28], $T_f$ = 4.6 K. For $R$ = Gd, $T_f$ is almost a factor of ten smaller than $\Theta$, consistent with values typical for strongly frustrated systems. The data for $R$ = Tb and Dy appear to be more complicated. For both samples, the separation of the FC and ZFC curves occurs at higher temperature than the peak in the ZFC magnetization. This observation is not unique to these binary quasicrystals, however, as observed previously in the Tb-Mg-Cd system [11] among others, and has been subject to several interpretations including the presence of a distribution of spin-freezing temperatures and/or the onset of short-range magnetic correlations above $T_f$ [10,11], and crystalline electric field (CEF) effects. The different behavior noted for i-Gd-Cd is of particular interest in this context since it is an *S*-state ion, which does not manifest any CEF anisotropy. Similar magnetization measurements on i-Ho-Cd, i-Er-Cd and i-Tm-Cd do not evidence FC/ZFC differences or signatures of magnetic ordering above $T = 2$ K, the lowest temperature that could be reached in the current measurements. We find the magnetic behavior of the i-*R*-Cd phase is in stark contrast to what is observed for the closely related $RCd_6$ approximants (see insets to Figure 4) where signatures that have been associated with long-range magnetic order are clearly seen [5-7].

Given the greater structural and chemical simplicity associated with a binary compound, the matched sets of i-*R*-Cd and $RCd_6$ for *R* = Gd – Tm form model systems that will allow us to determine, refine and test our understanding of the key features and properties associated with quasicrystalline structure and magnetism. Structurally, the i-*R*-Cd series presents the intriguing prospect of two very different quasicrystal compositions associated with the same approximant, possibly with very different *R* content in the "glue". Magnetically, the i-*R*-Cd series, coupled with its AFM-ordering $RCd_6$ approximants, allows for direct comparisons between the low-temperature magnetic states of crystalline and quasicrystalline phases with fundamentally similar *R*-based clusters. Our results, so far, support the idea that the presence of aperiodic, rather than periodic, order disrupts or frustrates long-range magnetic order.

**Methods**

Single grain quasicrystals were grown out of Cd–rich binary solutions. Typical procedure involved adding approximately 5 g of Cd (Alfa Aesar, 99.9999% purity) and 0.06 g of rare-earth elements (Ames Lab) into a 2 ml alumina crucible with a molar ratio of Cd:$R$ = 99.2:0.8 ($R$ = Y, Gd-Dy) or 99.4:0.6 ($R$ = Ho-Tm). The crucible with the starting elements was sealed in a fused silica ampoule under partial argon atmosphere, which was then heated up to 700°C, held at 700°C for 10 hours, cooled to 455°C in 3 hours and then slowly (roughly 2 °C/hour) cooled to 335°C, at which temperature the somewhat $R$-depleted, remaining Cd-rich solution was decanted with the assistance of a centrifuge. Clusters of single grains with a mass of about 0.2 g were obtained from each growth.

Elemental analysis was performed using wavelength-dispersive x-ray spectroscopy (WDS) in the electron probe microanalyzer of a JEOL JXA-8200 electron microprobe with a 20 kV beam voltage and a 5-µm spot size. For each sample, the measurement was done at 12 different locations on a polished surface. Differential Thermal Analysis (DTA) was performed on samples sealed in tantalum capsules under high-purity Argon using a Miyachi/Unitek pulsed laser system allowing welding of the capsule with little or no sample heating. Calorimetry was performed on a Netzsch 404C using baseline background subtraction and a heating rate of 10 K/minute.

Powder x-ray diffraction patterns were measured on a Rigaku Miniflex II desktop x-ray diffractometer (Cu $K_\alpha$ radiation). Samples were prepared by grinding single grains into powder, which were then mounted and measured on a Si single crystal zero-background sample holder. High-energy x-ray diffraction measurements were performed on station 6-ID-D at the Advanced Photon Source using 100 keV x-rays and an area detector (MAR 345) positioned 98 cm from the sample position. Data was taken in precession mode where the axis normal to the reciprocal lattice planes of interest is set at a small, but finite, half-cone angle with respect to the incident beam direction and rotated stepwise about it.

The dc magnetization measurements were performed in a Quantum Design Magnetic Property Measurement System (MPMS-5), superconducting quantum interference device (SQUID) magnetometer ($T$ = 1.8-350 K, $H_{max}$ = 5.5 T). Low field (50 Oe) magnetization was measured on warming for zero-field-cooled, and on cooling for field-cooled data.

# References

[*]goldman@ameslab.gov, [**]canfield@ameslab.gov

[1] Canfield, P. C., Fishing the Fermi Sea, *Nature Phys.* **4**, 167 (2008).

[2] Canfield, P. C., Caudle, M. L., Ho, C. -S., Kreyssig, A., Nandi, S., Kim, M. G., Lin, X., Kracher, A., Dennis, K. W., McCallum, R. W. & Goldman, A. I., Solution growth of a binary icosahedral quasicrystal of $Sc_{12}Zn_{88}$, *Phys. Rev. B* **81**, 020201(R) (2010).

[3] Tsai, A. P., Guo, J. Q. Abe, E., Takakura, H., & Sato, T. J., A stable binary quasicrystal, *Nature* **408**, 537-538 (2000).

[4] Takakura, H., Pay Gómez, C., Yamamoto, A., De Boissieu, M., & Tsai, A. P., Atomic structure of the binary icosahedral Yb-Cd quasicrystal, *Nature Mater.* **6**, 58-63 (2007).

[5] Tamura, R., Muro, Y., Hiroto, T., Nishimoto, K. & Takabatake, T., Long-range magnetic order in the quasicrystalline approximant $Cd_6Tb$, *Phys. Rev. B* **82**, 220201(R) (2010).

[6] Kim, M. G., Beutier, G., Kreyssig, A., Hiroto, T., Yamada, T., Kim, J. W., de Boissieu, M., Tamura, R., & Goldman, A. I., Antiferromagnetic order in the quasicrystal approximant $Cd_6Tb$ studied by x-ray resonant magnetic scattering, *Phys. Rev. B* **85**, 134442 (2012).

[7] Mori, A., Ota, H., Yoshiuchi, S., Iwakawa, K., Taga, Y., Hirose, Y., Takeuchi, T., Yamamoto, E., Haga, Y., Honda, F., Settai, R. & Ōnuki, Y., Electrical and Magnetic Properties of Quasicrystal Approximants $RCd_6$ (R: Rare Earth), *J. Phys. Soc. Jpn.* **81**, 024720 (2012).

[8] Fukamichi, K., *Physical Properties of Quasicrystals* (Springer-Verlag, Berlin, 1999), pp. 295–326.

[9] Hippert, F. & Prejean, J. J., Magnetism in Quasicrystals, *Phil. Mag.* **88**, 2175 (2008).

[10] Fisher, I. R., Cheon, K. O., Panchula, A. F., Canfield, P. C., Chernikov, M., Ott, H. R. & Dennis, K., Magnetic and transport properties of single-grain R-Mg-Zn icosahedral quasicrystals [R=Y, $Y_{1-x}Gd_x$, $Y_{1-x}Tb_x$, Tb, Dy, Ho, and Er], *Phys. Rev. B* **59**, 308 (1999).

[11] Sato, T. J., Short-range order and spin-glass-like freezing in *A*-Mg-R (*A* = Zn or Cd; R = rare-earth elements) magnetic quasicrystals, *Acta Cryst.* A**61**, 39 (2005).

[12] Sebastian, S. E., Huie, T., Fisher, I. R., Dennis, K. W. & Kramer, M. J., Magnetic properties of single grain R–Mg–Cd primitive icosahedral quasicrystals (R=Y, Gd, Tb or Dy), *Phil. Mag.* **84**, 1029 (2004).

[13] Jazbec, S., Jagličić, Z., Vrtnik, S., Wencka, M., Feuerbacher, M., Heggen, M., Roitsch, S. & Dolinšek, J., Geometric origin of magnetic frustration in the μ-Al$_4$Mn giant-unit-cell complex intermetallic, *J. Phys.: Condens. Matter* **23**, 045702 (2011).

[14] Wang, P., Stadnik, Z. M., Al-Qadi, K. & Przewoźnik, J., A comparative study of the magnetic properties of the 1/1 approximant Ag$_{50}$In$_{36}$Gd$_{14}$ and the icosahedral quasicrystal Ag$_{50}$In$_{36}$Gd$_{14}$, *J. Phys.: Condens. Matter* **21**, 436007 (2009).

[15] Ibuka, S., Iida, K. & Sato, T. J., Magnetic properties of the Ag–In–rare-earth 1/1 approximants, *J. Phys.: Condens. Matter* **23**, 056001 (2011).

[16] See, for example, Goldman, A. I. & Kelton, K. F., Quasicrystals and crystalline approximants, *Rev. Mod. Phys.* **65**, 213 (1993).

[17] Lifshitz, R., Symmetry of Magnetically Ordered Quasicrystals, *Phys. Rev. Lett.* **80**, 2717 (1998).

[18] Canfield, P. C., Solution Growth of Intermetallic Single Crystals: A Beginner's Guide, *Properties and Applications of Complex Intermetallics* (World Scientific, Singapore, 2010), pp. 93–111.

[19] Canfield, P. C. and Fisher, I. R., High-temperature solution growth of intermetallic single crystals and quasicrystals, *J. Cryst. Growth* **225**, 155 (2001).

[20] Gschneidner K. A. Jr., & Calderwood F. W., Cd-Gd (Cadmium-Gadolinium), Binary Alloy Phase Diagrams, II Ed., Ed. T.B. Massalski, Vol. 2, 1990, pp. 980-983.

[21] Kreyssig, A., Chang, S., Janssen, Y., Kim, J. W., Nandi, S., Yan, J. Q., Tan, L., McQueeney, R. J., Canfield, P. C. & Goldman, A. I., Crystallographic phase transition within the magnetically ordered state of Ce$_2$Fe$_{17}$, *Phys. Rev. B* **76**, 054421 (2007).

[22] Shannon, R. D., Revised Effective Ionic Radii and Systematic Study of Interatomic Distances in Halides and Chalcogenides, *Acta Cryst. A* **32**, 751 (1976).

[23] Larson, A. C. & Von Dreele, R. B., General Structure Analysis System (GSAS), *Los Alamos National Laboratory Report* LAUR 86-748, (2004).

[24] Pay Gómez, C. & Lidin, S., Comparative structural study of the disordered *M*Cd$_6$ quasicrystal approximants, *Phys. Rev. B* **68**, 024203 (2003).

[25] Kawana, D., Watanuki, T., Machida, A., Shobu, T., & Aoki, K., Intermediate-valence quasicrystal of a Cd-Yb alloy under pressure, *Phys. Rev. B* **81**, 220202(R) (2010).

[26] M. de Boissieu & A. P. Tsai, private communication.

[27] Janot, C., *Quasicrystals: A Primer* (Oxford University Press, New York, 1992).

[28] Mydosh, J. A., *Spin Glasses: An Experimental Introduction* (Taylor and Francis, London, 1993).

**Acknowledgments**

We acknowledge and thank W. Straszheim for the WDS measurements, D. S. Robinson for assistance with the high-energy x-ray diffraction measurements and R. J. McQueeney for useful discussions. The research was supported by the Office of the Basic Energy Sciences, Materials Sciences Division, U. S. Department of Energy (DOE). Ames Laboratory is operated for DOE by Iowa State University under contract No. DE-AC02-07CH11358. Use of the Advanced Photon Source was supported by the US DOE under Contract No. DE-AC02-06CH11357.

**Author Contributions**

AIG, PCC, SLB and AK designed the measurements; TK, SLB and PCC grew the samples, performed and analyzed the magnetization measurements; KWD performed the DTA measurements and analysis; TK, AJ, MR, AK and AIG performed the x-ray diffraction measurements and data analysis. AIG and PCC drafted the manuscript and all authors participated in the writing and review of the final draft.

**Competing financial interests**

The authors declare that they have no competing financial interests.

**Figure Legends**

**Figure 1.** <u>Revised Cadmium-Gadolinium binary phase diagram.</u>

The Cd-Gd binary phase diagram from Reference [20] is revised (shown in blue) to update the liquidus curve for low Gd concentrations and to include the quasicrystalline phase (labeled i-Gd-Cd). The liquidus curve was re-determined (solid circles) by decanting growths at different temperatures and weighing the crystalline ($GdCd_6$) or i-Gd-Cd products. Given the mass of starting material, mass of the yield and composition of the crystal, the molar percentage of Gd remaining in the liquid upon decant can be calculated. The new eutectic point is inferred to be <0.4 % of Gd. The central inset presents the differential scanning calorimetry data from single grains of the quasicrystalline phase taken on warming through their thermal decomposition at roughly 450 ˚C. The small feature near 320 ˚C is the melting of the small amount of residual Cd flux on the surfaces of the quasicrystal grains. The pentagonal dodecahedral morphology of the icosahedral phase is shown in the upper-left inset with a millimeter scale; grains as large as ~ 1.0 mm have been grown.

**Figure 2.** <u>Weiss temperature, $\Theta$, values versus de Gennes factor.</u>

Solid points present the Weiss temperature, $\Theta$, values of $R$-Cd quasicrystals (black circles) obtained from a linear fit of the high-temperature inverse magnetic susceptibility, $\chi^{-1}(T) = H/M(T)$, measured at either $H = 0.5$ or 1 Tesla (shown in the inset). Error bars are inferred from measurements on multiple samples and different fitting ranges. The red, blue and brown symbols represent the Weiss temperature, $\Theta$, values for other icosahedral quasicrystals: $R_8Mg_{42}Zn_{50}$ [10], $R_{10}Mg_{40}Cd_{50}$ [11] and $Gd_{14}Ag_{50}In_{36}$ [14].

**Figure 3.** <u>Single-grain and powder x-ray diffraction from i-Gd-Cd.</u>

High-energy x-ray diffraction patterns from a single grain of i-Gd-Cd were taken with the beam parallel to (a) the five-fold and (b) two-fold axes. The scaling of peak positions along the five-fold axis confirms that the i-Gd-Cd quasicrystals fall within the simple-icosahedral (P-type) structural family. The powder diffraction pattern from i-Gd-Cd is shown in panel (c) along with the indexing of prominent peaks. All peaks in the pattern belong either to the icosahedral phase or the residual Cd flux. The inset shows the scaling of the 6D quasilattice constants (determined from indexing the patterns) for $R$ = Gd – Tm as a function of ionic radius and compares the lattice constant determined from powder measurements of our $RCd_6$ samples with the values calculated from $a_{6D}$ as described in the text. Values for the ionic radii were taken from reference [22] for nine-fold coordination. For i-Y-Cd, the lattice constant for the approximant, derived from $a_{6D} = $ 7.955 Å (15.483 Å) and the measured value of $a = $ 15.482 Å are also in excellent agreement and lie between those values determined for $R$ = Dy and Tb. The powder diffraction patterns from the $RCd_6$ approximants were refined using the Rietveld package

GSAS [23] and the results are in good agreement with the published crystallographic data [24].

**Figure 4.** Low-temperature Field-Cooled (FC) and Zero-Field-Cooled (ZFC) magnetic susceptibility, $M(T)/H$, data.

Temperature-dependent FC and ZFC magnetization data measured at 50 Oe from (a) i-Gd-Cd, (b) i-Tb-Cd and (c) i-Dy-Cd. The insets compare the higher-field $M(T)/H$ data (ZFC) for icosahedral phase samples with their respective $R\text{Cd}_6$ approximants. The arrows in the insets indicate the location of clear magnetic ordering features for the $R\text{Cd}_6$ approximants [5-7] which are absent for the related i-$R$-Cd compounds (main figure).

| $R\mathrm{Cd}_x$ | $x$ (WDS) | $x$ (Magnetization) | $\Theta$ (K) | $a_{6D}$ (Å) |
|---|---|---|---|---|
| Gd | 7.88(18) | 7.98(7) | -41(1) | 7.972(4) |
| Tb | 7.69(17) | 7.89(7) | -21(1) | 7.958(4) |
| Dy | 7.50(9) | 7.51(6) | -11(1) | 7.949(5) |
| Ho | 7.60(13) | 7.80(9) | -6(1) | 7.935(5) |
| Er | 7.34(13) | 7.78(5) | -4(1) | 7.935(6) |
| Tm | 7.28(6) | 7.76(8) | -2(1) | 7.914(5) |
| Y | 7.48(16) | - | - | 7.955(5) |

**Table 1.** The composition of the icosahedral phase as inferred from wavelength-dispersive spectroscopy (WDS) and temperature-dependent magnetization (inset to Figure 2), the value of the Weiss temperature, $\Theta$, and the six-dimensional quasilattice constant, $a_{6D}$, determined by indexing the powder diffraction patterns from each sample [see inset to Figure 3(c)]. The WDS data were calibrated to the values measured for the respective $R\mathrm{Cd}_6$ samples we had grown. The $R$:Cd ratio was also extracted from the temperature dependent susceptibility by assuming that $R$ = Gd – Tm manifest full, trivalent, local moments and that the temperature-dependent magnetic susceptibility, $\chi(T) = M(T)/H$, could be fit to a Curie-Weiss law, $\chi(T) = C/(T-\Theta)$. For the heaviest rare-earth members of the i-$R$-Cd series, given the shrinking grain size, the $x$ value inferred from the magnetization data may become less reliable than the WDS value due to the increasing significance of small amount of residual Cd flux on the surfaces of the grains. For example, approximately 5% Cd (by mass) second phase on the i-Tm-Cd quasicrystal can shift the inferred value for $x$ from the WDS value of 7.28 to the magnetization derived value of 7.76. Errors in parenthesis for $x$ from the WDS measurements and $a_{6D}$ represent one standard deviation in the values. The error in $x$ from the magnetization measurements was estimated from fitting over different temperature ranges and weighing errors.

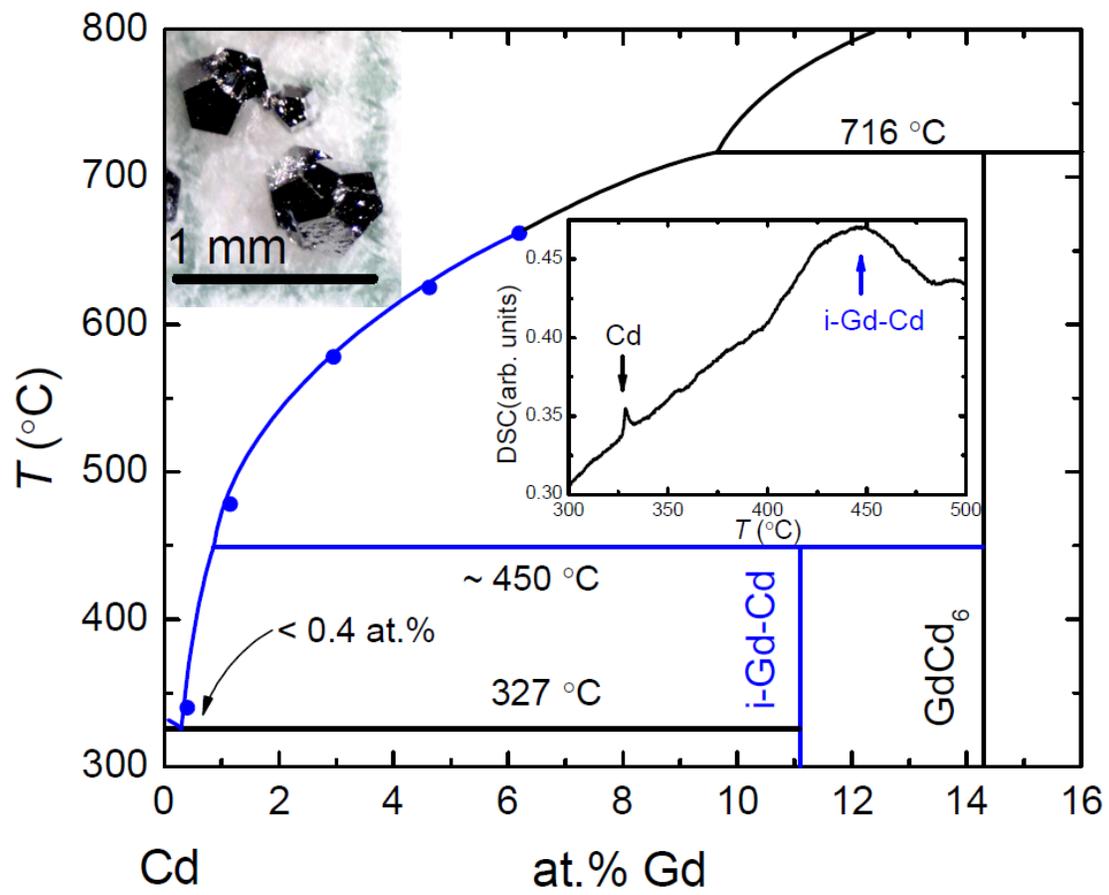

**Figure 1.**

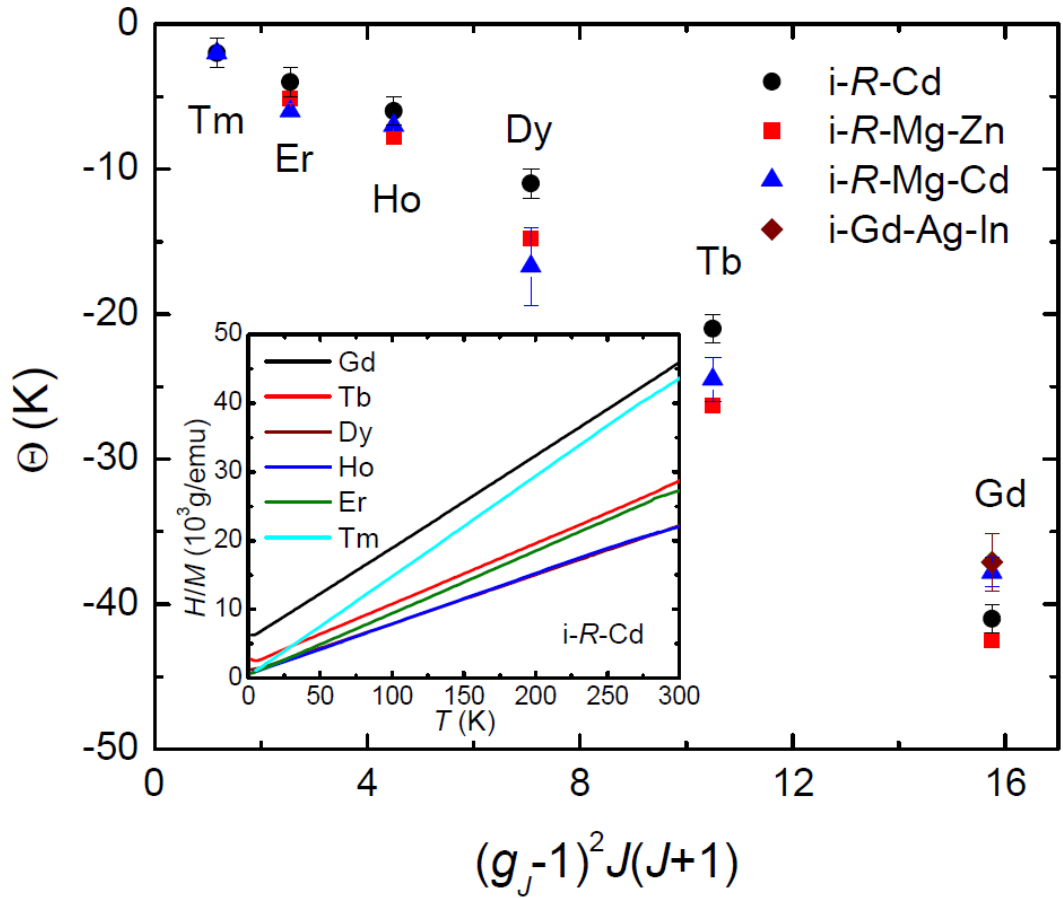

**Figure 2.**

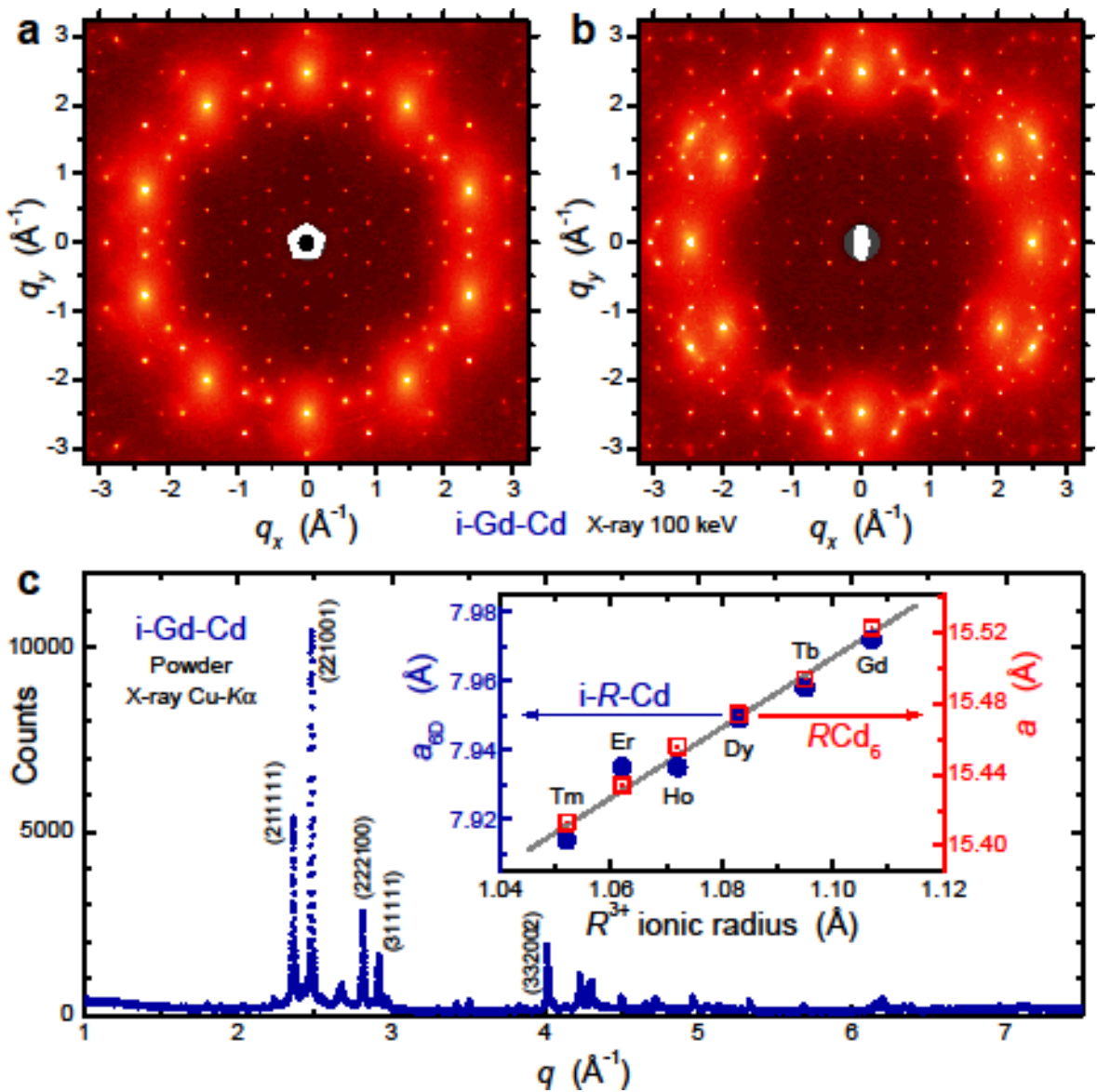

**Figure 3.**

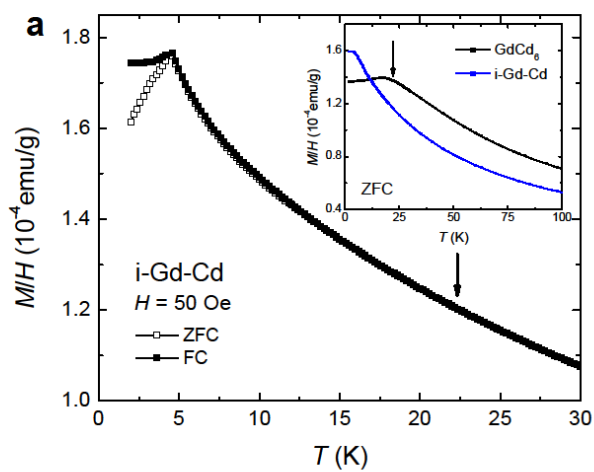

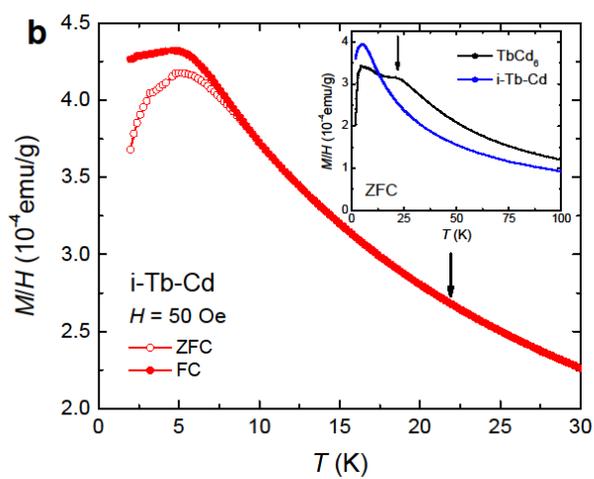

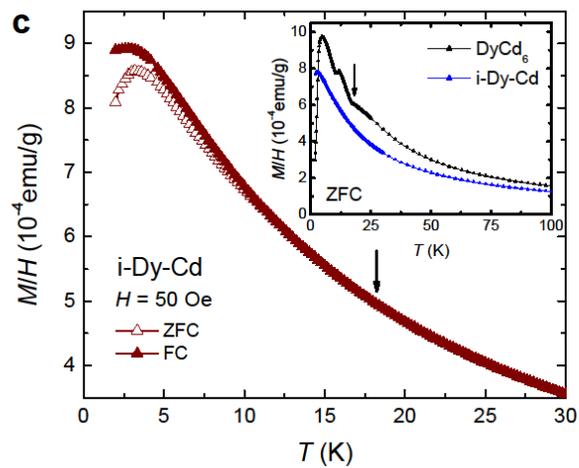

**Figure 4.**